\documentclass[letterpaper, 10 pt, conference]{IEEEtran}
\IEEEoverridecommandlockouts

\usepackage{cite}
\usepackage[pdftex]{graphicx}
\DeclareGraphicsExtensions{.pdf,.jpeg,.png}
\usepackage{empheq}
\usepackage{amssymb}
\usepackage{bbm}
\usepackage{upgreek}
\usepackage{todonotes}
\usepackage{xspace}
\usepackage{wasysym}
\usepackage{mathrsfs}
\usepackage{enumitem}
\usepackage{epstopdf}
\usepackage[overlay,absolute]{textpos}

\newcommand{\real}{\ensuremath{\mathbb{R}}}
\newcommand{\nat}{\ensuremath{\mathbb{N}}}

\newcommand{\R}{\ensuremath{{R}}}

\newcommand{\Au}{\ensuremath{\mathscr{A}}}
\newcommand{\Oh}{\ensuremath{\mathscr{O}}}

\newcommand{\smat}[1]{\ensuremath{\left[\begin{smallmatrix}#1\end{smallmatrix}\right]}}
\newcommand{\Hy}{\ensuremath{\mathcal{H}}}
\newcommand{\So}{\ensuremath{\mathcal{S}}}
\newcommand{\x}{\ensuremath{\mathsf{x}}}
\newcommand{\f}{\ensuremath{\mathsf{f}}}
\newcommand{\F}{\ensuremath{\mathsf{F}}}
\newcommand{\C}{\ensuremath{\mathsf{C}}}
\newcommand{\g}{\ensuremath{\mathsf{g}}}
\newcommand{\G}{\ensuremath{\mathsf{G}}}
\newcommand{\n}{\ensuremath{\mathsf{n}}}
\newcommand{\B}{\ensuremath{\mathsf{B}}}
\newcommand{\D}{\ensuremath{\mathsf{D}}}

\renewcommand{\R}{\ensuremath{\mathsf{R}}}
\newcommand{\CR}{\ensuremath{\overline{\C\backslash\R}}}
\newcommand{\DR}{\ensuremath{\overline{\D\backslash\R}}}

\newcommand{\Un}{\ensuremath{\mathsf{U}}}

\DeclareMathOperator{\dom}{dom}

\DeclareMathOperator{\gph}{gph}

\newtheorem{assumption}{Assumption}
\newtheorem{problem}{Problem}
\newtheorem{example}{Example}
\newtheorem{proposition}{Proposition}

\newtheorem{thm}{Theorem}
\newenvironment{theorem}{\begin{thm}}{\end{thm}}
\newtheorem{definition}{Definition}
\newenvironment{proof}{\noindent\textit{Proof.}}{\hfill $\square$}

\newtheorem{remark}{Remark}

\begin{document}

\title{A hybrid barrier certificate approach\\to satisfy linear temporal logic specifications}
\author{Andrea Bisoffi and Dimos V. Dimarogonas\thanks{
This work was supported in part by the European Research Council (ERC) through ERC StG BUCOPHSYS, the Swedish Research Council (VR), the Swedish Foundation for Strategic
Research (SSF), the Knut and Alice Wallenberg Foundation (KAW) and SRA ICT TNG project STaRT.
The authors are with the Department of Automatic Control,
School of Electrical Engineering, KTH Royal Institute of Technology, 100 44 Stockholm, Sweden. \small{\texttt{bisoffi@kth.se, dimos@kth.se}}}
}

\maketitle

\begin{abstract}
In this work we formulate the satisfaction of a (syntactically co-safe) linear temporal logic specification on a physical plant through a recent hybrid dynamical systems formalism. In order to solve this problem, we introduce an extension to such a hybrid system framework of the so-called eventuality property, which matches suitably the condition for the satisfaction of such a temporal logic specification. The eventuality property can be established through barrier certificates, which we derive for the considered hybrid system framework. Using a hybrid barrier certificate, we propose a solution to the original problem. Simulations illustrate the effectiveness of the proposed method.
\end{abstract}

\begin{textblock*}{\textwidth}(17mm,4mm)%
\small\bfseries{\color{red}\noindent\textcopyright \textcopyright 2018 IEEE. Personal use of this material is permitted. Permission from IEEE must be obtained for all other uses, in any current or future media, including reprinting/republishing this material for advertising or promotional purposes, creating new collective works, for resale or redistribution to servers or lists, or reuse of any copyrighted component of this work in other works.}
\end{textblock*}

\section{Introduction}
\label{sec:intro}

Linear Temporal Logic (LTL, see, e.g., \cite{baier2008principles,belta2017formal}) provides a tool to formulate richly expressive control specifications for continuous-time plants (e.g., high-level tasks for multi-robot systems). Since an LTL formula can be equivalently translated into an automaton \cite[Thm.~5.41]{baier2008principles}, the combination of the continuous-time plant and the automaton can be appealingly addressed through a hybrid system formalism  \cite{goebel2012hybrid}, in order to leverage available control tools for the continuous-time part. More precisely, we specify through the hybrid system how the solutions of the continuous-time plant generate a word of observations  (corresponding to regions of interest), following the terminology and approach of~\cite[Chap.~2]{belta2017formal}. We focus in this work on syntactically co-safe Linear Temporal Logic (sc-LTL), which is a relevant fragment of LTL. The fact that a word of observations satisfies the sc-LTL specification, corresponds to reaching a subset of the states of the automaton above after a \emph{finite} number of steps. Such a condition can be then conveniently encompassed into an \emph{eventuality} property of a suitable set of the whole state of the hybrid system.

The notion of eventuality is stated for continuous-time systems in~\cite[\S 3.2]{prajna2007convex}, and is paralleled here for a generic hybrid system~\cite{goebel2012hybrid} as the existence of a finite \emph{hybrid} time after which a given set is reached by all solutions (cf. Definition~\ref{def:event}). The eventuality property for such a hybrid system is an attractivity-like property with some distinct features. Indeed, it is a weaker property than finite time \emph{attractivity} \cite[Def.~3.1]{li2016results} 
because no settling-time function, independent of the considered solution, is required. The eventuality property bears similarities with the recurrence property in~\cite[\S 13.4.5]{petit2017feedback}, but the latter takes its full meaning for stochastic systems, as we argue more in detail in Remark~\ref{rem:recurrence}.
Barrier certificates to assess the eventuality property have been proposed again in~\cite{prajna2007convex} for a continuous-time setting. Certifying eventuality without explicitly computing solutions is the motivation behind barrier certificates for eventuality. 
We extend them in this work for the hybrid setting~\cite{goebel2012hybrid} as in Theorem~\ref{thm:barrierCert}.

The contributions of this paper are as follows. By formulating the satisfaction of a sc-LTL formula by a continuous-time plant as a hybrid system~\cite{goebel2012hybrid}, we are motivated to 
extend the eventuality property to such hybrid systems. We provide for them sufficient conditions of Lyapunov type in terms of barrier certificates, as a key contribution.
Finally, we provide a solution through barrier certificates to the problem of the satisfaction of a sc-LTL formula by a continuous-time plant.

We propose a hybrid barrier certificate approach to overcome the computational cost associated with discretizations into (possibly very large) finite transition systems of the continuous-time plant as in, e.g., \cite{tabuada2009verification,baier2008principles,belta2017formal}. For the same reasons, such a discretization is also avoided in, e.g., \cite{verdier2017formal,podelski2007region,dimitrova2014deductive}. As related work, the concept of eventuality agrees with the so-called \emph{region stability} of \cite[Def.~1]{podelski2007region}. To the best of the authors' knowledge, barrier certificates for eventuality have not been proposed for hybrid systems~\cite{goebel2012hybrid}, although barrier certificates for other properties were proposed for the hybrid automata described for instance in~\cite[\S 2]{prajna2004safety}. Such hybrid automata can be formulated in the formalism of \cite{goebel2012hybrid} as shown in~\cite[\S 1.4.1-1.4.2]{goebel2012hybrid}, but it is not possible to formulate as a hybrid automaton a \emph{generic} hybrid system \cite{goebel2012hybrid} (cf.~Equation~\eqref{eq:hsGen}), for which our main Theorem~\ref{thm:barrierCert} is derived. \cite{prajna2004safety}~uses barrier certificates for \emph{safety} on a hybrid automaton. \cite{podelski2007region}~proposes a method to enforce the \emph{region stability} above on a hybrid automaton, and is based on computing solutions, unlike a barrier certificate approach. \cite{dimitrova2014deductive}~proposes a proof system for alternating-time temporal logic {on a continuous-time system}. 
Finally, other works on barrier certificates for continuous-time systems (or on their counterparts for design, the so-called control barrier functions) are \cite{wisniewski2016converse,romdlony2016stabilization,wieland2007constructive,ames2016control} and references therein.

The structure of the paper is as follows. Section~\ref{sec:prel+probState} presents some preliminaries and the problem statement. Section~\ref{sec:barrier} defines the eventuality property and provides a barrier certificate for a generic hybrid system \cite{goebel2012hybrid} as a main result. Section~\ref{sec:scLTL} then applies such a tool to solve the considered problem. The solution is illustrated by a numerical example in Section~\ref{sec:sim}, and conclusions are in~Section~\ref{sec:concl}.
All the proofs are omitted due to space constraints.

\textit{Notation.} Given a set $S$, we denote its closure by $\overline{S}$ and its cardinality by $|S|$. $\nat$ is the set of the natural numbers. The logical operators \textit{not}, \textit{and}, \textit{or} are denoted by $\lnot$, $\wedge$, $\vee$. $\langle \cdot, \cdot \rangle$ defines the inner product between its two vector arguments. For a set-valued mapping $M\colon \real^n \rightrightarrows \real^n$, the domain of $M$ is $\dom M := \{ x \in \real^n  \colon M(x) \neq \emptyset \}$ and its graph is the set $\gph M:= \{(x,y)\in \real^n \times \real^n \colon y \in M(x)\}$. $\Oh(\cdot)$ denotes an asymptotic upper bound in algorithm analysis \cite[p.~47]{cormen2009introduction}.

\section{Preliminaries and problem statement}
\label{sec:prel+probState}

After some preliminaries about the two main ingredients of this work in Sections~\ref{sec:LTL} and \ref{sec:hs}, we can present the addressed problem in Section~\ref{sec:probStat}.

\subsection{Linear Temporal Logic and Finite State Automaton}
\label{sec:LTL}

This work is focused on the fragment of LTL called syntactically co-safe Linear Temporal Logic (sc-LTL), for whose definition we adopt the terminology and approach in~\cite[\S 2.1]{belta2017formal}. Since each sc-LTL formula $\psi$ 
can be translated into a Finite State Automaton (FSA) (as proven in, e.g., \cite[\S II.B]{bhatia2010sampling}), we consider in the sequel just the FSA representation of a sc-LTL formula as:

\begin{definition}\textit{ (Finite state automaton, semantics and acceptance condition~\cite[Def.~2.4]{belta2017formal})}
A \emph{finite state automaton} (FSA) is a tuple $\Au = (S, s_0, O, \delta, S_f)$, where: 
$S$ is a finite set of states,
$s_0 \in S$ is the initial state,
$O$ is a finite set of observations,
$\delta \colon S \times O \to S$ is a transition function\footnote{More precisely, $\delta$ is a \emph{partial} function: it does not map every element of its domain, i.e., it might not be defined for some $(s,o)$.}, $S_f \subset S$ is the set of accepting states. 
The \emph{semantics} of an FSA is defined over finite words of observations. For some $n\in \nat$, a run of $\Au$ over a finite word  of observations $w_O = w_O(1)w_O(2)\dots w_O(n)$ (with $w_O(k) \in O$ for all $k=1,\dots, n$) is a sequence $w_S = w_S(1)w_S(2)\dots w_S(n+1)$ where $w_S(1) = s_0$ and $w_S(k + 1) = \delta(w_S(k),w_O(k))\in S$ for all $k = 1, \dots, n$.  The word $w_O$ is \emph{accepted} by $\Au$ if the corresponding run $w_S$ ends in an accepting state of the automaton, i.e., $w_S(n + 1) \in S_f$.
\label{def:FSA}
\end{definition}

The set of the words accepted by $\Au$ coincides with the set of prefixes satisfying the corresponding sc-LTL formula $\psi$  \cite[p.~31]{belta2017formal}, so its satisfaction is guaranteed in a finite number of steps. Considering a deterministic FSA (due to the deterministic $\delta$ and a single $s_0$) is without loss of generality as a nondeterministic FSA can be translated into an equivalent deterministic FSA (see, e.g., \cite[Thm.~2.11]{hopcroft2006introduction}).

\begin{figure}
\centering
\includegraphics[width=0.45\textwidth]{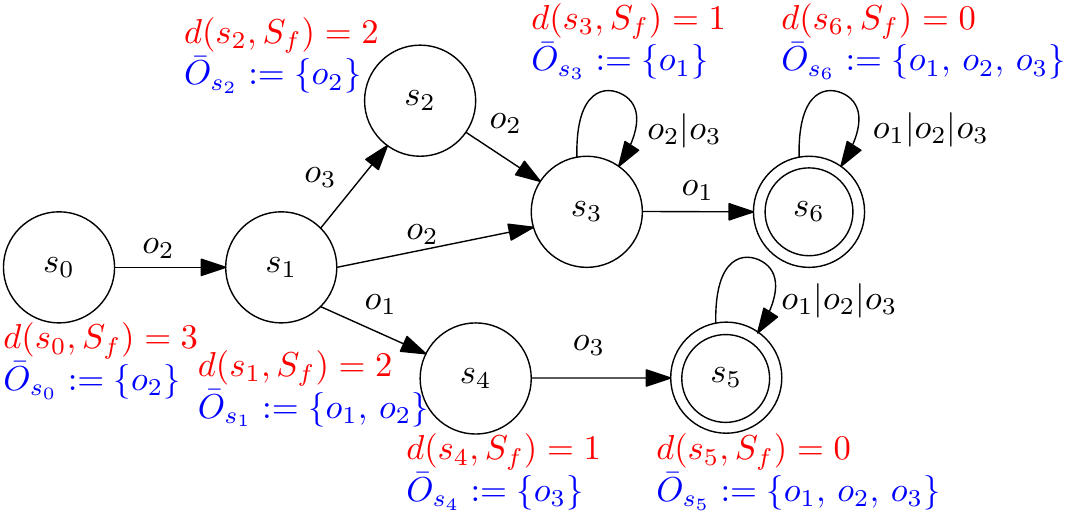}
\caption{The automaton corresponding to the formula in~\eqref{eq:scLTLformula}. The notation $o_i|\dots|o_j$ next to a transition means that such a transition is enabled if either $o_i$, \dots, or $o_j$ are generated. Double circles denote accepting states $S_f$. The meaning of red and blue labels for each $s_i$ is clarified in Section~\ref{sec:sim}.}
\label{fig:automaton}
\end{figure}

We present now an example of how a sc-LTL formula is translated in a standard way into an FSA. 
\begin{example}
\label{ex:automaton}
For $O=\{o_1,\,o_2,\,o_3\}$, let us consider the following sc-LTL formula
\begin{equation}
\label{eq:scLTLformula}
o_2 \wedge \bigcirc \Big(
\big( (\lnot o_1 \Un o_2) \wedge (\lozenge o_1) \big) \vee \big( o_1 \wedge (\bigcirc o_3) \big)
\Big)
\end{equation}
where the symbols $\bigcirc$, $\Un$, $\lozenge$ denote respectively the temporal logic operators \textit{next}, \textit{until}, \textit{eventually} as in \cite[p.~28]{belta2017formal}. Then, an intuitive rendering of the fact that a word $w_O$ satisfies the formula in~\eqref{eq:scLTLformula}, is as follows:\newline
\textit{$o_2$ is present as first element of $w_O$; next we have that: \textit{(i)} $o_2$ is present at some point in $w_O$ and until then $o_1$ is not present, and $o_1$ is also eventually present, \emph{or} \textit{(ii)} $o_1$ is present as first element of $w_O$ and $o_3$ as next one.}\newline
The automaton corresponding to the formula in~\eqref{eq:scLTLformula} is in Figure~\ref{fig:automaton}. It has been obtained through the tool \textsc{ltl2ba} \cite{gastin2001fast}, partially simplified as in~\cite[Ex.~2.8]{belta2017formal} (because of the definition of $\delta\colon S \times O \to S$ in Definition~\ref{def:FSA}, transitions can be triggered by at most one observation, and not by multiple observations), and made deterministic as indicated above.
\end{example}

\subsection{Hybrid dynamical systems}
\label{sec:hs}

In order to model the evolution in continuous time for the plant and the one corresponding to discrete updates in the logical state of the FSA $\Au$, we consider the hybrid dynamical system $\Hy$ \cite{goebel2012hybrid} with state $\x\in \real^\n$:
\begin{subequations}
\label{eq:hsGen}
\begin{empheq}[left={\Hy:\empheqlbrace}]{align}
& \dot \x  \in \F(\x), \quad \x \in \C \label{eq:hsGenFlow} \\
& \x^+  \in \G(\x), \quad \x \in \D. \label{eq:hsGenJump}
\end{empheq}
\end{subequations}
The state $\x$ is allowed to evolve according to the \textit{flow map} given by the differential inclusion $\F$ (that can be reduced to a differential equation) when it belongs to the \textit{flow set} $\C$ and according to the \textit{jump map} given by the difference inclusion $\G$ (that can be reduced to a difference equation) when it belongs to the \textit{jump set} $\D$. A solution $\phi$ to~\eqref{eq:hsGen} \cite[Def.~2.6]{goebel2012hybrid} is then parametrized naturally by two time directions as $(t,j)\mapsto \phi(t,j)$, where $t$ denotes the continuous time and $j$ acts as a counter of the jumps occurred. The subset of $\real^2$ of points $(t,j)$ where a solution $\phi$ is defined is called a hybrid time domain \cite[Def.~2.3]{goebel2012hybrid} and is denoted by $\dom \phi$. 
We will refer the reader to specific points in~\cite{goebel2012hybrid} whenever further details are needed.

\subsection{Problem statement}
\label{sec:probStat}

Based on Sections~\ref{sec:LTL} and \ref{sec:hs}, we introduce in this section the system in which we are interested, and describe then our problem statement.

A sc-LTL formula $\psi$ is given in the form of a finite state automaton $\Au := (S,s_0,O,\delta,S_f)$ from Definition~\ref{def:FSA} as discussed in Section~\ref{sec:LTL} and we assume in this work that such sc-LTL formula can indeed be satisfied:
\begin{assumption}
\label{ass:feas}
For the FSA $\Au := (S,s_0,O,\delta,S_f)$, there exists $s_f \in S_f$ that is reachable from the initial state $s_0$. Moreover, \emph{without loss of generality}, we remove from $\Au$ all the states that are not reachable from $s_0$ and from which no accepting state $s_f$ can be reached.
\end{assumption}

The approach of, e.g., \cite{belta2017formal} associates an observation with the states of a transition system to be controlled, and the possible words of observations generated by this transition system are checked against the sc-LTL formula to find those satisfying the formula. Instead of the transition system, we consider here directly the continuous-time dynamics described by a linear time invariant plant with state $\xi\in \real^n$ and control $u\in \real^m$
\begin{equation}
\label{eq:LTIplant}
\dot \xi = A \xi + B u,
\end{equation}
and we specify how to associate each solution to~\eqref{eq:LTIplant} with a word of observations, which should conform to the sc-LTL formula. Specifically, each observation $o$ corresponds to a region of interest $D_o$ for the state $\xi$ in~\eqref{eq:LTIplant}, such that
 \begin{equation}
\label{eq:Do}
D_o \subset \real^n \text{ is a \emph{compact} set with nonempty interior}.
\end{equation}
Indeed, solutions to the hybrid system
\begin{equation}
\label{eq:hsJustPlant}
\begin{aligned}
& \dot \xi  = A \xi + B u, & & \xi \in \overline{\real^n \backslash D_o}\\
& \xi^+ = \xi, & & \xi \in D_o
\end{aligned}
\end{equation}
are enforced to flow under a suitable action for $u$, which is specified in Section~\ref{sec:scLTL}, and when they jump from $D_o$, we say that the solution has generated the observation $o$. In order that the words of observations arising from such jumps of the solutions conform to the sc-LTL formula given by $\Au$, we further constraint the evolution of~\eqref{eq:hsJustPlant} as:
\begin{subequations}
\label{eq:hsWithInps}
\begin{align}
\left.
\begin{aligned}
\dot s & = 0\\
\dot o & = 0\\
\dot \xi & = A \xi + B u
\end{aligned}
\right\}
&
(s,o,\xi)\in C\label{eq:hsWithInpsFlow}\\
\left.
\begin{aligned}
s^+ & = \delta(s,o)\\
o^+ & = \bar o\\
\xi^+ & = \xi
\end{aligned}
\right\}
&
(s,o,\xi)\in D. \label{eq:hsWithInpsJump}
\end{align}
In~\eqref{eq:hsWithInps}, $s$ and $o$ do not change during flow. Corresponding to a jump, the current logical state $s\in S$ of $\Au$ in Definition~\ref{def:FSA} is updated to $s^+$ through the transition function $\delta$ of $\Au$ based on the observation $o$ generated by the solution. The observation $o^+$ we \emph{want} the solution to generate next in order to conform to the sc-LTL formula, is updated according to the discrete-time input $\bar o$. Due to such associated decision, $\bar o$ parallels the continuous-time input $u$ and is also specified in Section~\ref{sec:scLTL}. $\xi$ does not change across jumps. Finally, to specify $C$ and $D$ in~\eqref{eq:hsWithInpsFlow}-\eqref{eq:hsWithInpsJump}, define for each $s\in S$
\begin{equation}
\label{eq:Os}
O_s := \{ o \in O \colon \delta(s,o) \text{ is defined}\}.
\end{equation}
The overall flow and jump sets are then
\begin{align}
C:= \{(s,o,\xi) \colon s\in S,\, & o\in O_s,\, \xi \in \overline{\real^n \backslash D_o} \} \label{eq:hsWithInpsFlowSet}\\
D:= \{(s,o,\xi) \colon s\in S,\, & o\in O_s,\, \xi \in D_o\} \label{eq:hsWithInpsJumpSet},
\end{align}
so that jumps are allowed only in the set $D$ comprising \emph{all possible} $s\in S$, $o\in O_s$ as defined in~\eqref{eq:Os} and $\xi \in D_o$, whereas for all such $s$ and $o$, solutions can only flow before they reach $D_o$.
\end{subequations}
\begin{remark}
Since the set of observations $O$ is independent of the state set $S$ of the FSA $\Au$ in Definition~\ref{def:FSA}, we emphasize that $D_o$ is determined only by $o$ (and \emph{not} by $s$). Solutions to~\eqref{eq:hsWithInps} are allowed to jump only after they reach $D_o$, although they can flow through $D_{o'}$ with $o' \neq o$.
\end{remark}

As in~\cite{sanfelice2013existence}, $u$ and $\bar o$ play the role of hybrid inputs. By acting on $u$ and $\bar o$, then, \eqref{eq:hsWithInps} should generate through jumps a word of observations that is accepted by the sc-LTL. So, the input $\bar o$ in~\eqref{eq:hsWithInpsJump} needs to be constrained for a given $s^+$ as
\begin{equation}
\label{eq:barO}
\bar o \in O_{s^+},
\end{equation}
where from its definition in~\eqref{eq:Os}, $O_{s^+}$ contains only those elements $\bar o'$ for which $\delta(s^+,\bar o')$ is defined. 

Given the constraint~\eqref{eq:barO} for~\eqref{eq:hsWithInps}, the satisfaction of the sc-LTL formula is then equivalent, based on Definition~\ref{def:FSA}, to the solution property that the component $s$ of the solution to~\eqref{eq:hsWithInps} at hybrid time $(T,J)$ satisfies $s(T,J)\in S_f$ for some finite $T\ge 0$ and $J \ge 0$. We then have:
\begin{problem}
\label{probl:state}
For system~\eqref{eq:hsWithInps} under the constraint~\eqref{eq:barO}, find a control law for $u$ and $\bar o$ such that for some finite $T \ge 0$ and $J\ge 0$, $s(T,J)\in S_f$.
\end{problem}

To solve Problem~\ref{probl:state} and guarantee the eventuality property of solutions, we develop sufficient conditions in terms of barrier certificates in the sense of~\cite{prajna2007convex} for hybrid systems \cite{goebel2012hybrid} in Section~\ref{sec:barrier}. In Section~\ref{sec:scLTL}, we propose such a barrier certificate for Problem~\ref{probl:state} after we specified a (possible) control law for $u$ and $\bar o$.

\section{Eventuality property for hybrid systems through barrier certificates}
\label{sec:barrier}

In the scope of this section we consider the generic hybrid system $\Hy$ in~\eqref{eq:hsGen} with state $\x \in \real^\n$ and data $(\F, \C, \G, \D)$.

We require that \eqref{eq:hsGen} satisfies mild regularity assumptions as in \cite[Ass.~6.5]{goebel2012hybrid}\footnote{Broadly speaking, \cite[Ass.~6.5]{goebel2012hybrid} guarantees that stability properties are uniform and robust w.r.t. small perturbations (see \cite[pp.~139, 169]{goebel2012hybrid}).} together with a so-called \emph{viability condition} so that basic existence of solutions is guaranteed, as in the following Assumption~\ref{ass:hbc}. $T_\C(x)$ below denotes the tangent cone to the set $\C$ at a point $\x$ as in~\cite[Def.~5.12 and Fig.~5.4]{goebel2012hybrid}.
\begin{assumption}
\label{ass:hbc}
(1) The data $(\F,\C,\G,\D)$ satisfy the hybrid basic conditions as in~\cite[Ass.~6.5]{goebel2012hybrid}, that is: $\C$ and $\D$ are closed sets in $\real^\n$; the set-valued mappings $\F$ and $\G$ have a closed graph and are locally bounded relative to $\C$ and $\D$, respectively; $\C \subset \dom F$ and $\D \subset \dom G$; $\F(\x)$ is convex for each $\x \in \C$. (2) For every $\upxi \in \C\backslash\D$ there exists a neighborhood $\mathsf{U}$ of $\upxi$ such that for every $\x \in \mathsf{U} \cap \C$, $\F(\x) \cap T_\mathsf{C}(\x) \neq \emptyset$.
\end{assumption}

Motivated by \cite[Thm.~3.5]{prajna2007convex}, we generalize the eventuality property for a hybrid system $\Hy$ in~\eqref{eq:hsGen}, as in the following Definition~\ref{def:event}. $\So_\Hy$ below denotes the set of all maximal solutions $\phi$ to $\Hy$ as in \cite[p.~33]{goebel2012hybrid}, and 
a solution is said to be \emph{maximal} if it cannot be extended, as per~\cite[Def.~2.7]{goebel2012hybrid}.
\begin{definition}\textit{ (Eventuality property w.r.t. a set $\R$)}
\label{def:event}
Consider $\Hy$ in~\eqref{eq:hsGen} and a closed set $\R\subset \C \cup \D$. The eventuality property w.r.t. the set $\R$ holds if for \emph{each} solution $\phi \in \So_\Hy$ there exist finite $T\ge 0$ and $J\ge 0$ such that $\phi(T,J)\in \R$ and for all $(t,j)\in \dom \phi$ with $t+j < T+J$, $\phi(t,j) \in \C \cup \D$.
\end{definition}

Due to the nonuniqueness of solutions  inherent in $\Hy$ in~\eqref{eq:hsGen},
we require that the eventuality property is satisfied by \emph{all} solutions. Nonuniqueness is also motivated by the fact that the vector field in~\cite[Thm.~3.5]{prajna2007convex} is assumed to be only continuous.
\begin{remark}
\label{rem:recurrence}
The eventuality property in Definition~\ref{def:event} bears similarities with the recurrence property in~\cite[\S 13.4.5]{petit2017feedback}, which is however fully meaningful for \emph{stochastic} hybrid dynamical systems. Moreover, recurrence excludes finite escape times altogether, whereas here they are admitted if the solution reaches $\R$ before escaping to infinity.
\end{remark}

Inspired by \cite[Thm.~3.5 and Remark~3.6]{prajna2007convex}, we derive sufficient conditions to guarantee eventuality for $\Hy$ in~\eqref{eq:hsGen}.
\begin{theorem}[Barrier certificate for eventuality]
\label{thm:barrierCert}
Consider $\Hy$ in~\eqref{eq:hsGen} satisfying Assumption~\ref{ass:hbc}. Let $\R\subset \C \cup \D$ be a closed set such that
\begin{subequations}
\label{eq:Bcond}
\begin{equation}
\G(\DR) \subset \C \cup \D. \label{eq:Bcond_jumpout}
\end{equation}
If there exist a function $\B$, continuous on $\CR \cup \DR$ and differentiable on an open neighborhood of $\CR$, and $\epsilon >0$ such that:
\begin{align}
& \B \text{ is bounded from below on } \CR \cup \DR \label{eq:Bcond_boundedbelow}\\
& \B \text{ is radially unbounded\footnotemark~on } \CR \cup \DR \label{eq:Bcond_radiallyUnbounded}\\
& \langle \nabla \B(\x), \f \rangle < -\epsilon \quad \forall \x \in \CR,\, \forall \f \in \F(\x) \label{eq:Bcond_flow}\\
& \B(\g) - \B(\x) < -\epsilon \quad \forall \x \in \DR,\, \forall \g \in \G(\x), \label{eq:Bcond_jump}
\end{align}
\end{subequations}
\footnotetext{Equivalently: if the sequence of points $\{ \x_i \}_{i=1}^{+ \infty}$ is unbounded and $\x_i \in \CR \cup \DR$ for all $i$, then also the sequence $\{ \B(\x_i) \}_{i=1}^{+ \infty}$ is unbounded. \label{footnote:radUnb}}%
then the eventuality property w.r.t. the set $\R$ holds, as in Definition~\ref{def:event}. We call $\B$ a \emph{barrier certificate} w.r.t. $\R$.
\end{theorem}

Some comments are in order.
Condition~\eqref{eq:Bcond_jumpout} is a \emph{necessary} condition to have the eventuality property in Definition~\ref{def:event}, which involves all maximal solutions to $\Hy$. Indeed, if \eqref{eq:Bcond_jumpout} did not hold, some solutions could jump out of $\C \cup \D$ without ever being in $\R$. \eqref{eq:Bcond_boundedbelow}, \eqref{eq:Bcond_flow}, \eqref{eq:Bcond_jump} guarantee essentially that the decrease of $\B$ will eventually lead any solution $\phi$ to reach $\R$. \eqref{eq:Bcond_radiallyUnbounded} excludes the existence of solutions with finite escape time that grow unbounded \emph{without} reaching $\R$.

\section{A barrier certificate for sc-LTL satisfaction}
\label{sec:scLTL}

In this section, we first specify the control law for $u$ and $\bar o$ for~\eqref{eq:hsWithInps}, and then propose a barrier certificate as in Theorem~\ref{thm:barrierCert}, which guarantees that the eventuality property w.r.t. a suitable set, as required in Problem~\ref{probl:state}, is achieved.

To this end, we introduce a shortest-path distance notion for the FSA $\Au$ of Definition~\ref{def:FSA} that was embedded in the hybrid system~\eqref{eq:hsWithInps} as explained in Section~\ref{sec:probStat}.
The FSA $\Au$ in Definition~\ref{def:FSA} can be seen as a digraph where each $s$ represents a vertex, and each observation $o\in O_s$ in~\eqref{eq:Os} labels an edge from $s$ to $\delta(s,o)$. We compute then for each node $s\in S$ its shortest-path distance $\hat d$ to any other node $s_f\in S_f$ as
\begin{subequations}
\label{eq:distOnAutomaton}
\begin{equation}
\label{eq:hatDist}
\begin{aligned}
(&s,s_f)\mapsto \hat d(s,s_f)\\
& := \begin{cases}
\infty, \text{\qquad\quad if there is no path from $s$ to $s_f$}\\
\begin{minipage}{3.8cm}
minimum number of edges \\
in any path from $s$ to $s_f$
\end{minipage},  \qquad \text{otherwise,}
\end{cases}
\end{aligned}
\end{equation}
based on the breadth-first search algorithm \cite[\S 22.2]{cormen2009introduction}. Through~\eqref{eq:hatDist}, we define the distance of $s \in S$ to the set of accepting states $S_f$ as
\begin{equation}
\label{eq:dist}
d(s,S_f) := \min_{s_f \in S_f} \hat d(s,s_f),
\end{equation}
\end{subequations}
which is the minimum shortest-path distance from $s$ over the accepting states $s_f\in S_f$. At a jump, after which the logical state takes the next value $s^+$, the discrete-time input $\bar o$ is selected, consistently with~\eqref{eq:barO}, as \emph{any} element in $O_{s^+}$
which additionally makes the distance to $S_f$ decrease strictly:
\begin{multline}
\label{eq:barOsel}
\bar o \in \bar O_{s^+} := \{ \bar o' \in O_{s^+} \colon \\
d(\delta(s^+,\bar o'),S_f) < d(s^+,S_f) \text{ if } s^+ \notin S_f \} ,
\end{multline}
where $d(s^+,S_f)$ is the distance to $S_f$ just after the jump and $d(\delta(s^+,\bar o'),S_f)$ is the distance to $S_f$ that is achieved when selecting $\bar o'$ as the next observation we want to generate. Showing the nonemptiness of $\bar O_{s^+}$ for each $s^+ \notin S_f$ is part of the proof of Proposition~\ref{prop:barrier_scLTL} below.

We turn then to the selection of the continuous-time input $u$. To keep the exposition focused on the core contribution of this work, we ask to control the \emph{whole state} to a generic setpoint $\bar \xi \in \real^n$, as in the following Assumption~\ref{ass:contr}. This assumption can be relaxed if we only want to control some \emph{output} to a setpoint (see \cite[\S 23.6]{hespanha09}).
\begin{assumption}
\label{ass:contr}
The pair $(A,B)$ is controllable and such that for all $\bar \xi \in \real^n$, there exists a unique solution $\bar u$ to
\begin{equation*}
\label{eq:setPointControl}
A \bar\xi + B \bar u = 0.
\end{equation*}
\end{assumption}

Under Assumption~\ref{ass:contr}, for every $o\in O$ it is possible \cite[p.~116]{hespanha09} to find Lyapunov functions $U_o$ certifying asymptotic stability of a generic point $c_o$ belonging to the \emph{interior} of $D_o$ in~\eqref{eq:Do} as
\begin{equation}
\label{eq:Uo}
U_o(\xi) := (\xi-c_o)^T P_o (\xi-c_o)
\end{equation}
where $P_o$ is positive definite. Based on these Lyapunov functions, it is possible in turn to select $u$ as
\begin{equation}
\label{eq:uSel}
u = u_o -K_o (\xi - c_o)
\end{equation}
with $u_o$ defined such that $A c_o + B u_o = 0$ and, for instance, $K_o:= \tfrac{1}{2} B^T P_o$ \cite[\S 12.4]{hespanha09}. Then,
\begin{multline}
\label{eq:lyapEqPoQo}
(\xi-c_o)^T [ P_o (A- B K_o) + (A- B K_o)^T P_o] (\xi-c_o)\\
=-(\xi-c_o)^T Q_o (\xi-c_o) < 0\quad \forall \xi \neq c_o
\end{multline}
where the positive definiteness of $Q_o$ is guaranteed by the definition of $K_o$ and the controllability of the pair $(A,B)$ in Assumption~\ref{ass:contr}.

With the input selections in~\eqref{eq:barOsel} and \eqref{eq:uSel}, \eqref{eq:hsWithInps} becomes
\begin{subequations}
\label{eq:hsCL}
\begin{align}
&
\begin{bmatrix}
\dot s\\
\dot o\\
\dot \xi\\
\end{bmatrix}
=
\begin{bmatrix}
0\\
0\\
(A - B K_o) (\xi - c_o)\\
\end{bmatrix}
=: f(x),\, 
(s,o,\xi)\in C \label{eq:hsCLflow}\\
&
\begin{bmatrix}
s^+ \\
o^+ \\
\xi^+\\
\end{bmatrix}
\in
\begin{bmatrix}
\delta(s,o)\\
\bar O_{\delta(s,o)}\\
\xi \\
\end{bmatrix}
=:G(x),\,
(s,o,\xi)\in D, \label{eq:hsCLjump}
\end{align}
\end{subequations}
where the total state is defined concisely as
\begin{equation}
\label{eq:overallState}
x := (s,o,\xi).
\end{equation}
Consistently with the FSA $\Au$ and the policy~\eqref{eq:barOsel}, the initial conditions for the the logical states are selected as:
\begin{equation}
\label{eq:hsCLic}
s(0,0)=s_0, \quad o(0,0) \in \bar O_{s_0}	.
\end{equation}

To solve Problem~\ref{probl:state}, we propose then a barrier certificate for eventuality in the sense of Theorem~\ref{thm:barrierCert}. For the closed set
\begin{equation}
\label{eq:hsR}
R :=\{ (s,o,\xi) \colon s\in S_f,\,o\in O_s,\, \xi \in \real^n\} ,
\end{equation}
the barrier certificate w.r.t. $R$ is, for the state $x$ in~\eqref{eq:overallState},
\begin{equation}
\label{eq:barrier_scLTL}
B(x) := d(s,S_f) + \lambda U_o(\xi),
\end{equation}
where $d$ is in~\eqref{eq:dist}, $U_o$ is in~\eqref{eq:Uo} and $\lambda>0$ is a sufficiently small parameter
whose existence is part of the proof of Proposition~\ref{prop:barrier_scLTL} below, which solves Problem~\ref{probl:state}.
Note that an energy function equal to the distance $d$ was already used in~\cite[Def.~3.3]{ding2014ltl}.
\begin{proposition}
\label{prop:barrier_scLTL}
Under Assumptions~\ref{ass:feas} and \ref{ass:contr}, there exist a sufficiently small $\lambda>0$ such that $B$ in~\eqref{eq:barrier_scLTL} is a barrier certificate w.r.t. the set $R$ in~\eqref{eq:hsR}, in the sense of Theorem~\ref{thm:barrierCert}, for the hybrid system in~\eqref{eq:hsCL}.
\end{proposition}

\begin{remark}
Our solution can be compared to the automata-based approach in~\cite[\S 5.3]{belta2017formal}. We share with that approach the computational cost to translate the sc-LTL formula into the FSA $\Au$. Our approach requires computing the matrices $K_o$'s and the distances $d$ in~\eqref{eq:distOnAutomaton} (based on the distances $\hat d$ for each accepting state $s_f$), but \emph{not} $\lambda$. Therefore, the overall cost can be shown to be $\Oh(|O| n^3 + |S_f|(v_\Au + e_\Au))$ where $v_\Au$ and $e_\Au$ are respectively the number of vertices and edges of $\Au$. On the other hand, we do not build any product automaton of $\Au$ and the transition system discretizing the continuous-time plant, so we do not have the cost $\Oh(|S_p| |\Sigma|)$ of~\cite[\S 5.3]{belta2017formal} where $|S_p|$ and $|\Sigma|$ are respectively the cardinalities of the set of states of such product automaton and of the set of inputs of such transition system. In order that the transition system captures well the plant, $|S_p|$ easily becomes very large.
\end{remark}

\begin{figure}[!t]
\centering
\includegraphics[width=0.49\textwidth]{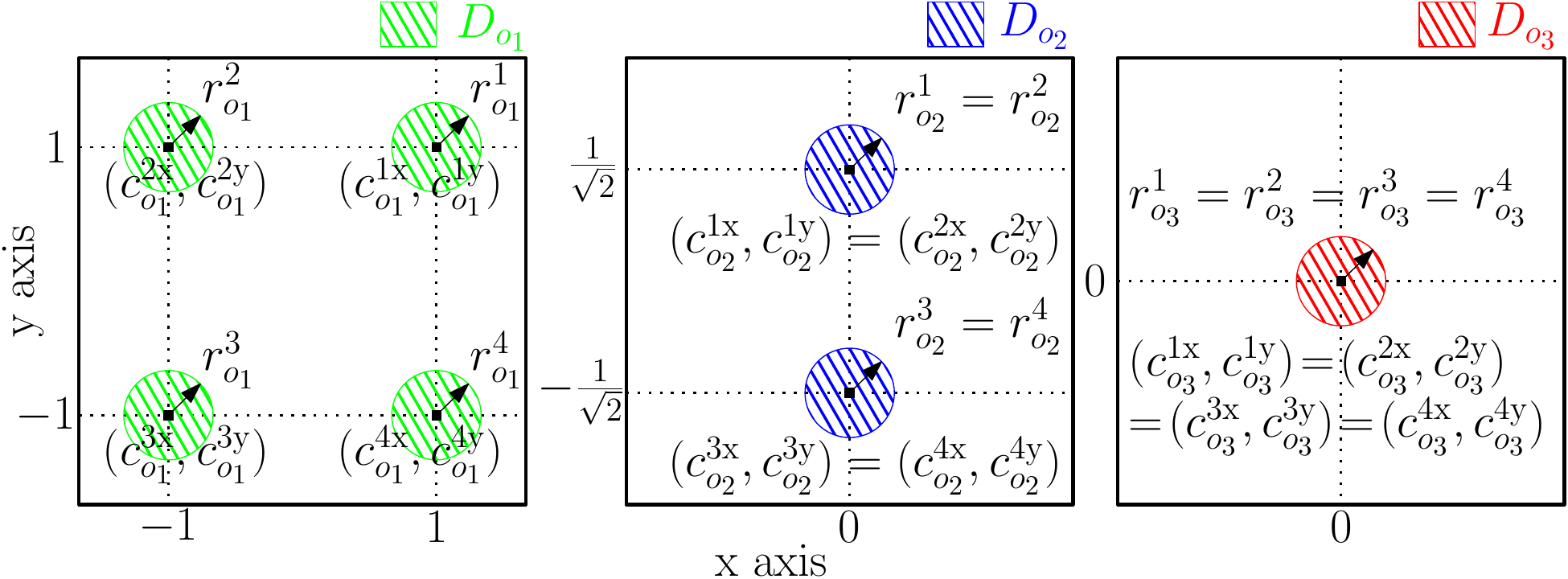}
\caption{The sets $D_{o_k}$ in~\eqref{eq:Dok} for $k=1,\,2,\,3$ projected for each agent $i=1,\dots,4$ onto the x-y plane and the corresponding selected parameters $c^{i \mathrm{x}}_{o_k}$, $c^{i \mathrm{y}}_{o_k}$, $r^i_{o_k}$ ($r^i_{o_k}=0.1$ for all $i$, $k$).}
\label{fig:Do}
\end{figure}

\section{Simulations}
\label{sec:sim}

In this section we present a numerical example to illustrate the eventuality property w.r.t. $R$ in~\eqref{eq:hsR} of the solutions to~\eqref{eq:hsCL} when a barrier certificate $B$ is found as in~\eqref{eq:barrier_scLTL}, that is, in the setting of Proposition~\ref{prop:barrier_scLTL}.

The sc-LTL specification is given by the formula in~\eqref{eq:scLTLformula}, so this section continues Example~\ref{ex:automaton}. We have then
\begin{equation}
\label{eq:SOF}
\! S:=\{ s_0, \dots, s_6 \}, \, O:=\{ o_1, o_2, o_3 \},\, S_f:=\{s_5,s_6\}.
\end{equation}
We assign to each vertex of the corresponding automaton in Figure~\ref{fig:automaton} a distance $d$ as in~\eqref{eq:distOnAutomaton} and report it in Figure~\ref{fig:automaton} (red labels).
For each $s\in S$, we compute the set $\bar O_s$ based on~\eqref{eq:barOsel} and report it in Figure~\ref{fig:automaton} (blue labels). This specifies the jump map in~\eqref{eq:hsCLjump}.

As for the flow map in~\eqref{eq:hsCLflow}, consider a continuous state $\xi$ that is the stack of the $\mathrm{x}$ and $\mathrm{y}$ position of 4 agents:
\begin{equation}
\label{eq:xiStack}
\xi:=(\xi^{1\mathrm{x}},\xi^{1\mathrm{y}},\xi^{2\mathrm{x}},\xi^{2\mathrm{y}},\xi^{3\mathrm{x}},\xi^{3\mathrm{y}},\xi^{4\mathrm{x}},\xi^{4\mathrm{y}}).
\end{equation}
The matrices $A$ and $B$ are constructed starting from a Laplacian matrix $L$ corresponding to the agent connections:
\begin{equation}
A:= - L \otimes I_2, \, B := I_8, \, L := \smat{1 & -1 & 0 & 0\\ -1 & 3 & -1 & -1\\ 0 & -1 & 1 & 0\\ 0 & -1 & 0 & 1\\}
\end{equation}
where $\otimes$ denotes the Kronecker product and $I_m$ an $m \times m$ identity matrix. With these $A$ and $B$, the state $\xi$ and input $u$ can be interpreted as a group of follower and leader agents, respectively (see~\cite[\S 10.6]{mesbahi2010graph}). Moreover, $A$ and $B$ satisfy Assumption~\ref{ass:contr}.
The matrix $P_o$ in \eqref{eq:Uo} is taken for all $o$'s as the unique solution $P$ of the Lyapunov equation
\begin{subequations}
\label{eq:lyapEq+Ko=K}
\begin{equation}
\label{eq:lyapEq}
P A + A^T P - P B B^T P = - 2 P,
\end{equation}
and take for each $o$
\begin{equation}
\label{eq:Ko=K}
K_o:=K:= \tfrac{1}{2} B^T P.
\end{equation}
\end{subequations}
\eqref{eq:lyapEq+Ko=K} verifies \eqref{eq:lyapEqPoQo} by simple computations as in~\cite[\S 12.4]{hespanha09}. Define then the components of $c_o$ by analogy with~\eqref{eq:xiStack} as
\begin{equation*}
c_o:=(c_o^{1\mathrm{x}},c_o^{1\mathrm{y}},c_o^{2\mathrm{x}},c_o^{2\mathrm{y}},c_o^{3\mathrm{x}},
c_o^{3\mathrm{y}},c_o^{4\mathrm{x}},c_o^{4\mathrm{y}}).
\end{equation*}

Consider the set $D_{o_k}$ ($k=1,\,2,\,3$) in Figure~\ref{fig:Do} (complying with~\eqref{eq:Do}) that can be described analytically as 
\begin{equation}
\label{eq:Dok}
D_{o_k} := \{ x \colon (\xi^{i\mathrm{x}}-c^{i \mathrm{x}}_{o_k})^2 + (\xi^{i\mathrm{y}}-c^{i \mathrm{y}}_{o_k})^2 \le (r^i_{o_k})^2, \, i = 1, \dots, 4 \}
\end{equation}
where $r^{i}_{o_k}$ is the radius corresponding to the observation $o_k$ for the agent $i$. See Figure~\ref{fig:Do}. Based on~\eqref{eq:Dok}, the overall flow and jump sets $C$ and $D$ in~\eqref{eq:hsCL} are obtained from~\eqref{eq:hsWithInpsFlowSet}-\eqref{eq:hsWithInpsJumpSet}.

\begin{figure}[!t]
\centering
\includegraphics[scale=.8]{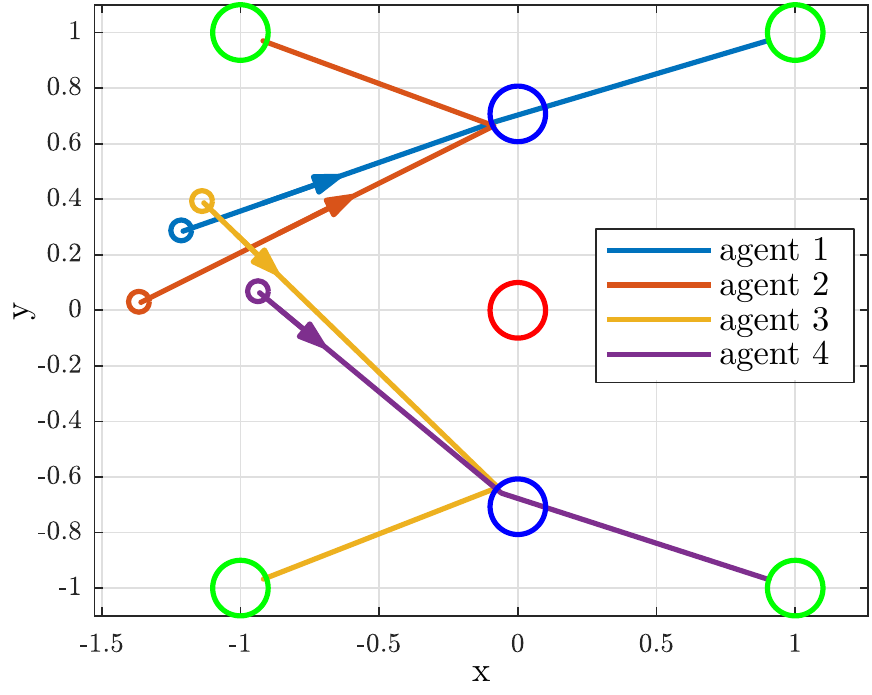}\\
\smallskip
\includegraphics[scale=.8]{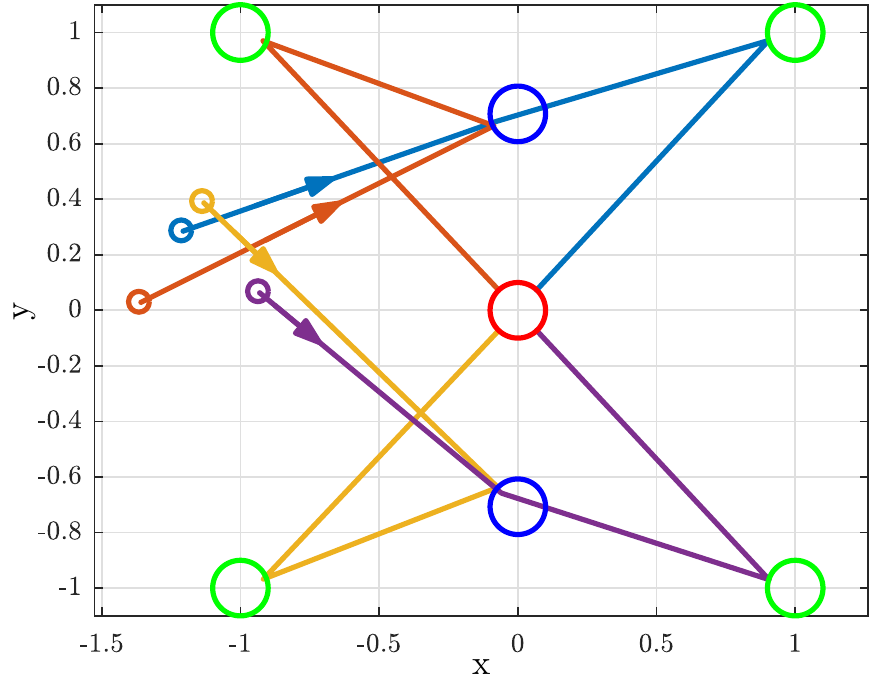}
\caption{Evolution of the x and y components of the agents, where the initial condition is denoted by a small circle. The big circles represent the sets $D_{o_k}$ corresponding to the generation of observations (cf. Figure~\ref{fig:Do}). The top (bottom, respectively) part shows a solution generating the observations $o_2$, $o_2$, $o_3$ ($o_2$, $o_1$, $o_3$) corresponding to the sequence $s_0$, $s_1$, $s_3$, $s_6$ ($s_0$, $s_1$, $s_4$, $s_5$) eventually reaching an accepting state of the automaton in Figure~\ref{fig:automaton}.}
\label{fig:path}
\end{figure}

\begin{figure}[!t]
\centering
\includegraphics[scale=.85]{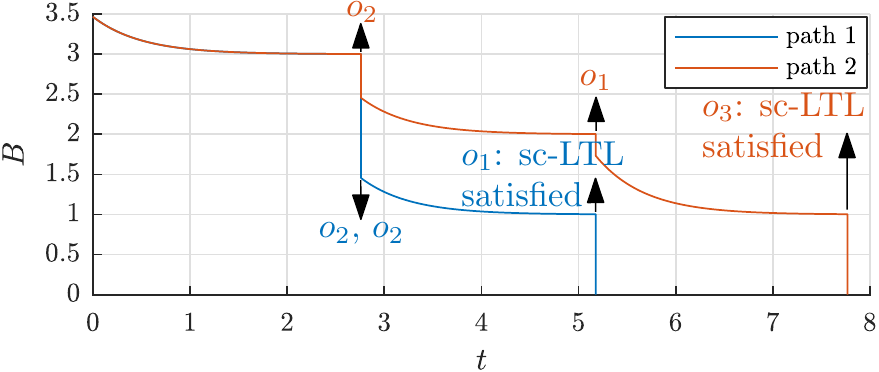}
\caption{Evolution of the barrier certificate $B$ in~\eqref{eq:barrier_sim} along the two solutions in Figure~\ref{fig:path}. The vertical arrows indicate the generated observations along the two paths up to satisfaction of the sc-LTL formula. The evolution of $B$ is truncated when solutions reach $R$ in~\eqref{eq:hsR}.}
\label{fig:barrier}
\end{figure}

From the automaton in Figure~\ref{fig:automaton}, the sets $D_{o_k}$ in~\eqref{eq:Dok} and $P$ in~\eqref{eq:lyapEq}, a sufficiently small $\lambda=0.0172$ can be found, as guaranteed by Proposition~\ref{prop:barrier_scLTL}. We emphasize that $\lambda$ is computed for  illustrating $B$, but is \emph{not} needed to implement our control strategy. With such $\lambda$ and the distances $d$ reported in Figure~\ref{fig:automaton}, the barrier certificate in~\eqref{eq:barrier_scLTL} becomes then:
\begin{equation}
\label{eq:barrier_sim}
B(x):= d(s,S_f)+ \lambda (\xi - c_o)^T P (\xi - c_o).
\end{equation}
This barrier certificate guarantees that all solutions to~\eqref{eq:hsCL} satisfy the eventuality property with respect to $R$ in~\eqref{eq:hsR}, and in particular with respect to the accepting state set $S_f$ in~\eqref{eq:SOF} of the FSA $\Au$ corresponding to the sc-LTL formula in~\eqref{eq:scLTLformula}. Indeed, two solutions satisfying the eventuality property are depicted in Figure~\ref{fig:path} and correspond to the sequence $s_0$, $s_1$, $s_3$, $s_6$ (path~1, top of Figure~\ref{fig:path}) and $s_0$, $s_1$, $s_4$, $s_5$ (path~2, bottom of Figure~\ref{fig:path}), which arise from the two admissible control sequences for $\bar o$ selected based on the $\bar O_s$ reported in Figure~\ref{fig:automaton}. For path~1, for instance, we note that the agents move so that the observations $o_2$, $o_2$ and then $o_1$ are generated (cf. Figure~\ref{fig:Do} for the sets $D_{o_k}$), thereby reaching the accepting state $s_6$ within $R$. 
The evolution of the barrier certificate along solutions for path~1 and 2 is shown in Figure~\ref{fig:barrier}, together with the generated observations that lead to satisfaction of the sc-LTL formula. The associated strict decrease from~\eqref{eq:Bcond_flow} and \eqref{eq:Bcond_jump} is evident. $B$ certifies eventuality for multiple paths that equally lead to satisfaction of the sc-LTL formula, thus generalizing the sequential setting of~\cite{guinaldo2017hybrid}.


\section{Conclusions and future developments}
\label{sec:concl}

In this work we have extended the eventuality property in~\cite{prajna2007convex} and the associated barrier certificates to the hybrid setting~\cite{goebel2012hybrid}. The resulting hybrid barrier certificate provides a solution to the problem of the satisfaction of a sc-LTL specification by a continuous-time linear time-invariant physical system.

\bibliographystyle{IEEEtran}
\bibliography{refs,IEEEabrv}

\end{document}